\documentclass[a4paper]{article}
\usepackage[T1]{fontenc}
\usepackage{hyperref}
\usepackage{amsmath}
\usepackage{amssymb}
\usepackage{amsthm}
\usepackage{graphicx}
\usepackage[rgb]{xcolor}
\usepackage[margin=3cm]{geometry}
\usepackage{tikz}

\newtheorem{theorem}{Theorem}
\newtheorem{proposition}{Proposition}
\newcommand{\pf}{\noindent {\it Proof.} \,}

\begin{document}

\title{
Random geometric graphs in high dimension 
}

\author{
    Vittorio Erba$^{1,2,a}$, Sebastiano Ariosto$^{1}$, Marco Gherardi$^{1,2}$ and Pietro Rotondo$^{2}$
\\[0.5cm]
\normalsize $^{1}$~Dipartimento di Fisica dell'Universit\`a di Milano,\\
\normalsize $^{2}$~Istituto Nazionale di Fisica Nucleare, sezione di Milano,\\
\footnotesize Via Celoria 16, 20100 Milano, Italy.\\
\\
\normalsize $^{a}$ Corresponding author. Email: vittorio.erba@unimi.it
\\
\\
\normalsize doi: 10.1103/PhysRevE.102.012306
}

\date{\today}

\maketitle
\begin{abstract} 
\noindent
Many machine learning algorithms used for dimensional reduction and manifold learning leverage on the computation of the nearest neighbours to each point of a dataset to perform their tasks.
These proximity relations define a so-called geometric graph, where two nodes are linked if they are sufficiently close to each other. 
\emph{Random geometric graphs}, where the positions of nodes are randomly generated in a subset of $\mathbb{R}^{d}$, offer a null model to study typical properties of datasets and of machine learning algorithms.
Up to now, most of the literature focused on the characterization of low-dimensional random geometric graphs whereas typical datasets of interest in machine learning live in high-dimensional spaces ($d \gg 10^{2}$).
In this work, we consider the infinite dimensions limit of hard and soft random geometric graphs and we show how to compute the average number of subgraphs of given finite size $k$, e.g.~the average number of $k$-cliques. This analysis highlights that local observables display different behaviors depending on the chosen ensemble: soft random geometric graphs with continuous activation functions converge to the naive infinite dimensional limit provided by Erd\"os-R\'enyi graphs, whereas hard random geometric graphs can show systematic deviations from it.   
We present numerical evidence that our analytical results, exact in infinite dimensions, provide a good approximation also for dimension $d\gtrsim10$. 
\end{abstract}

\section{Introduction}
Random geometric graphs (RGGs) are networks whose nodes are $d$-dimensional randomly generated vectors from some probability distribution over $\mathbb{R}^d$, and edges link nodes only if their distance does not exceed a threshold distance $r$ \cite{penrose2003RandomGeometricGraphs}.
As such, their connectivity structure encodes information about the spatial structure of the nodes, and on the space they are embedded in: for this reason they are widely used in modeling complex systems in which geometric constraints play a fundamental role, such as 
transport \cite{barthelemy2011SpatialNetworks,PhysRevE.96.032316},
wireless \cite{giles2015BetweennessCentralityDense}, and social networks \cite{Nettleton:CSR:2013,Bonato:proc:2010}. 

Most of the results on RGGs have been established in the low-dimensional regime $d\leq3$ \cite{penrose2003RandomGeometricGraphs, barthelemy2011SpatialNetworks, estrada2015RandomRectangularGraphs, allen-perkins2018RandomSphericalGraphs, penrose2016ConnectivitySoftRandom, PhysRevE.91.042136}.
However, the high-dimensional limit $d\rightarrow\infty$ has recently gathered interest. Indeed in the era of big data and machine learning, typical datasets are made of vectors of hundreds of components (think for instance to the workhorse model in computer vision, the MNIST dataset of handrwritten digits); understanding how high-dimensional geometry works, and how it affects the proximity structure of datasets is crucial for the correct usage of manifold learning algorithms (from dimensional reduction protocols \cite{Tenenbaum:Science:2000, Roweis:Science:2000} to  intrinsic dimension estimators \cite{erba2019IntrinsicDimensionEstimation}), and for the creation of novel procedures tailored for the high-dimensional regime with benefits for dimensional reduction and clustering algorithms.
With this idea in mind, high-dimensional RGGs become a perfect null model for unstructured data, to benchmark and compare against real world datasets \cite{gorban2018blessing, bobrowski2018TopologyRandomGeometric}.

On the more mathematical side, it is an open problem to understand whether high-dimensional RGGs converge (as a statistical \emph{ensemble}) to Erd\"os-R\'enyi graphs; rigorous results for RGGs with nodes uniformly distributed on the sphere can be found in \cite{devroye2011HighDimensionalRandomGeometric,bubeck2016TestingHighDimensional,avrachenkov2020CliquesHighDimensionalRandom} and suggest that high-$d$ RGGs are similar to Erd\"os-R\'enyi graphs. On the other hand, the clustering coefficient of RGGs with nodes uniformly distributed on the hypercube shows systematic deviations from the ERG prediction \cite{dall2002RandomGeometricGraphs}. 
A related but different question is whether the critical behaviour of high dimensional RGGs converges to that of Erd\"os-R\'enyi graphs: see \cite{heydenreich2019lace} for more information.

In this work, we present a general framework for the computation of the average value of local observables of high-dimensional hard and soft RGGs. To this end, we exploit a multivariate version of the central limit theorem (CLT) to show that the joint probability of rescaled distances between nodes is normal-distributed, and we compute and characterize its correlation matrix.

We evaluate the average number of $M$-cliques, i.e. of fully-connected subgraphs with $M$ vertices, in high-dimensional RGGs.
We point out that these local observables show systematic deviations from the ERG prediction in hard-RGGs (whenever the hypothesis of the CLT are satisfied), whereas we observe convergence to ERG for soft-RGGs with continuous activation functions.   
This implies that the form of the activation function as well as the probability distribution on the nodes are crucial elements in studying the convergence of RGGs to ERGs.

Finally, we present numerical evidence that our analytical results do not hold only for $d\rightarrow\infty$, but provide a good approximation even in finite dimensions as low as $d\sim10$. This suggests that the high-dimensional limit of RGGs could be seen as a 0-th order term of a series expansion in $d$, possibly giving perturbative access to analytical results for low-dimensional RGGs.

In summary, the main results of our manuscript are the following:
\begin{itemize}
    \item[(i)] we systematically establish (under hypotheses on node positions resembling those of the CLT) when we should expect deviations from the ERG prediction in the infinite dimensional limit, by studying the behavior of the average number of cliques for hard and soft RGGs. It is worth remarking that observing this deviation for $k$-cliques is a strong indicator that most of the other subgraphs will display systematic deviations from the naive ERG prediction as well;
    \item[(ii)] in the case where the average number of cliques does not converge to the ERG prediction (i.e., for hard RGGs), we provide a quantitative analysis, based on the multivariate CLT, that well reproduces the non-trivial limit behavior of the properties considered;
    \item[(iii)] we numerically show that the high-dimensional approximation under which we derive our results gives accurate results even in moderately low dimension $d \sim 10$.
\end{itemize}

The manuscript is organized as follows.
In Section~\ref{sec:def} we introduce the notation and define the \emph{ensembles} of RGGs that we will study.
In Section~\ref{sec:CLT} we use a multivariate version of the central limit theorem to derive an explicit expression for the joint probability distribution of the distances of $M$ randomly drawn vectors in the limit of high dimension. This will be the crucial tool to compute averages of observables in high-dimensional RGGs.
Finally, in Section~\ref{sec:res}, we present our results on the average number of $M$-cliques for hard and soft RGGs alongside with numerical simulations.

\section{\emph{Hard} and \emph{soft} random geometric graphs}
\label{sec:def}

\textbf{Note on terminology:} in the literature, random geometric graphs are those with hard activation function (see later in this Section). Here, when omitting the adjectives "hard" or "soft" we generically refer to both.
\\

\begin{figure}
    \centering    
    \scalebox{0.9}{%
        \includegraphics{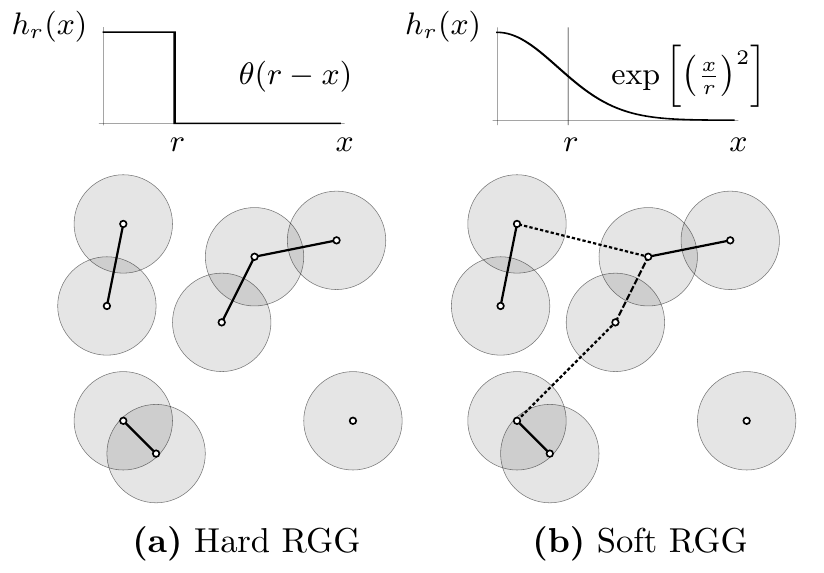}
    }
    \caption{\footnotesize \textbf{Example of hard and soft random geometric graphs.}
        Small circles denote nodes embedded in $\mathbb{R}^{2}$ drawn randomly with the uniform measure on $[0,1]^{2}$ and shaded circles highlight a region of radius $r/2$ around nodes.
        Solid lines highlight the actual edges of the represented graphs.
        On the top of the graph representations, the activation function used to build them are displayed.
        \textbf{(a)} In a hard random geometric graph at cutoff $r$, the only selected edges are those with nodes closer than $r$ (in the picture, the nodes whose shaded regions intersect). 
        \textbf{(b)} In a soft random geometric graph, edges are selected based on a continuous activation function $h_{r}(x)$. 
        If two nodes are at distance $d$ between each other, then the edge that connects them will be chosen with probability $h_{r}(d)$.
        In the picture, dotted edges are those edges that have been chosen by the soft random geometric graphs even though the distance between nodes was larger than $r$. 
        Vice versa, dashed edges are those at distance smaller than $r$, but not selected in that specific instance of the soft random geometric graph.
    }
    \label{fig:RGG}
\end{figure}

A \emph{random geometric graph} is a graph whose nodes are random points in $\mathbb{R}^{d}$, and whose edges are randomly generated based on the mutual distances between the nodes (see Figure~\ref{fig:RGG}).
Let us be more precise, starting by nodes.
We consider a probability distribution $\nu$ over $\mathbb{R}^{d}$, and we draw $N$ i.i.d. samples $\{\vec{x}_{i}\}_{i=1}^{N}$ from $\nu$; these will be the nodes of the random geometric graph.
Among the possible choices of $\nu$, a very common one is the uniform distribution on the $d$-dimensional hypercube $[0,1]^{d}$, i.e. 
\begin{equation}    
    \nu^{\rm cube}(\vec{x}) = \prod_{k=1}^{d} \theta(x^{k})\theta(1-x^{k})
\end{equation}
where $\theta$ is the Heaviside theta, and superscripts denote coordinates.
We will consider more in general probability distributions $\nu$ that are factorized and equally distributed over the coordinates, i.e.
\begin{equation} \label{eq:factorized} 
    \nu(\vec{x}) = \prod_{k=1}^{d} \tau(x^{k})
\end{equation}
where $\tau$ is a probability distribution on $\mathbb{R}$ with finite first and second moments.
In this case, the coordinates of all nodes $\{x_{i}^{k}\}$, with $1\leq i \leq N$ and $1\leq k \leq d$ are i.i.d. random variables with law $\tau$.

Now, for each pair of nodes $x,y$ we compute the distance $d(x,y)$ and we add the link $e=(x,y)$ to the edge set of the random geometric graph with probability $h(d(x,y))$, where $h:\mathbb{R}^{+}\rightarrow[0,1]$ is the so-called \emph{activation function} of the random geometric graph.
The activation function describes how likely it is for two nodes to be linked based on their distance, and will typically be a monotone decreasing function, with the idea that closer nodes will be linked with higher probability than further ones; we will consider monotone decreasing activation functions, with $h(0)=1$ and $h(+\infty)=0$.

Usually, the activation function is labeled by a parameter $r\in\mathbb{R}^{+}$ that describes the \emph{typical} distance at which a pair of nodes will be considered close enough to be linked with a non-trivial probability, for example $h_{r}(r)=\frac{1}{2}$. 
In this case, the statistical properties of random geometric graphs can be investigated as functions of $r$.

In this work, we will consider two types of activation functions.
The first one is that of \emph{hard} random geometric graphs, i.e.
\begin{equation}
    \begin{split}
        h^{\rm hard}_{r} (x) = \theta(r-x)\, .
    \end{split}
\end{equation}
In this case, all pairs of nodes with distance smaller than $r$ will be deterministically linked by an edge.
The second one is that of \emph{soft} random geometric graphs (also called random connection models in the literature), i.e. those with $h_{r}(x)$ at least continuous in $x$. 
A common choice in the literature (see for example \cite{giles2015BetweennessCentralityDense,kartun-giles2019ShapeShortestPaths}) is to employ the so-called \emph{Reyleigh fading} activation functions, i.e.
\begin{equation}
    \begin{split}
        h^{\rm rayleigh}_{r} (x) = \exp\left[ -\xi \left(\frac{x}{r}\right)^{\eta} \right]  \, ,
    \end{split}
\end{equation}
where $\xi = \log(2)$ guarantees that $h_{r}(r)=\frac{1}{2}$.

The last ingredient to be discussed is the distance function $d(x,y)$.
We will consider the $p$-norms $\mathbb{R}^{d}$
\begin{equation}
    \begin{split}
        ||\vec{x}||_{p} = \sqrt[p]{ \sum_{i=1}^{d} | x^{k} |^{p} } \, .
    \end{split}
\end{equation}
Notice that $p$-norms are norms only for $p\geq1$, as for $0<p<1$ the triangle inequality is not satisfied.
It this case, one can show that $||\vec{x}-\vec{y}||_{p}^{p}$ defines nonetheless a distance.
Thus, we will define and consider the distances
\begin{equation}
    \begin{split}
        d_{p}(\vec{x},\vec{y}) = ||\vec{x}-\vec{y}||_{p}^{\min\left(1,p\right)}
        \, .
    \end{split}
\end{equation}

\section{A central limit theorem for distances in high dimension}
\label{sec:CLT}

As a first step in our analysis, we are interested in computing the high-dimensional limit of the joint probability distribution of the distances between $M$ random points $\{\vec{x}_{i}\}_{i=1}^{M} \subset \mathbb{R}^{d}$, drawn independently from the factorized distribution $\nu$ in Equation~\eqref{eq:factorized}:
\begin{equation}\label{eq:PIdist}
    \begin{split}
        &\Pi(d_{(1,2)},d_{(1,3)},\dots d_{(M-1,M)}) \\ 
        &\qquad=
        \int \prod_{i=1}^{M} \nu(\vec{x}_{i}) dx_{i} \prod_{1\leq i < j \leq M} \delta \left( d_{p}(\vec{x}_{i},\vec{x}_{j}) -d_{(i,j)} \right)
        \, .
    \end{split}
\end{equation}

Since the distance $d_{p}(\vec{x},\vec{y})$ between two vectors $\vec{x},\vec{y}$ is a function of the sum of $d$ i.i.d. random variables, we expect that for $d \rightarrow \infty$ it converges to its average value $d \mu$ by the law of large numbers. Correspondingly, let us define the rescaled variables  
\begin{equation} \label{eq:defQ}
    \begin{split}
        q_{(i,j)} &= \frac{\left[d_{p}(\vec{x_{i}},\vec{x_{j}}) \right]^{\max(1,p)} - d \, \mu}{\sqrt{d}} \\
                  &= \frac{1}{\sqrt{d}} \sum_{k=1}^{d} \left( |x_{i}^{k} -x_{j}^{k}|^{p} - \mu \right) \\
                  &= \frac{1}{\sqrt{d}} \sum_{k=1}^{d} q_{(i,j)}^{k}\, .
    \end{split}
\end{equation}
where 
\begin{equation}
    \mu = \int dx dy \, \tau(x)\tau(y) | x - y |^{p} \, .
\end{equation}
Notice that the random vectors $\boldsymbol{q}_{k}=(q_{(1,2)}^{k},q_{(1,3)}^{k},\dots q_{(M-1,M)}^{k}) \in \mathbb{R}^{\binom{M}{2}}$, with $1\leq k\leq d$, are statistically independent and identically distributed, and that by definition the expected value of $\boldsymbol{q}_{k}$ is the null vector.
Notice also that the components of the vectors $\boldsymbol{q}_{k}$ are naturally indexed by lexicographically ordered multi-indices, as they are related to the distances between pairs of points along the $k$-th dimension; to distinguish such vectors from the Euclidean ones, we type them in boldface.

The vector $\boldsymbol{q} = (q_{(1,2)},q_{(1,3)},\dots q_{(M-1,M)})$ is a sum of i.i.d. multivariate random variables, and satisfies the following central limit theorem:
\begin{theorem}[Multivariate central limit theorem]\label{thm:CLT}
Let $\boldsymbol{q}^{1}, \boldsymbol{q}^{2} \dots \boldsymbol{q}^{d}$ be i.i.d. random vectors in $\mathbb{R}^{\binom{M}{2}}$ with null mean and covariance matrix $\boldsymbol{\Sigma}_{(i,j),(k,l)} = \mathbb{E}\left[ q^{1}_{(i,j)} q^{1}_{(k,l)} \right] $. Then
    \begin{equation}
        \begin{split}
            \boldsymbol{q}=\frac{1}{\sqrt{d}} \sum_{k=1}^{d} \boldsymbol{q}^{k} \, 
        \end{split}
    \end{equation}
is Gaussian-distributed with null mean and covariance $\boldsymbol\Sigma$ in the limit $d\rightarrow\infty$. 
\end{theorem}

The general formal proof can be found in \cite{van2000asymptotic} (see Proposition 2.17).  
A "physicist" approach to the proof would be to compute the characteristic function of $\boldsymbol{q}$ and to expand it to the leading order for large $d$.
It is worth noticing that the first neglected term in the expansion is of relative order $1/\sqrt{d}$, and may depend on $M$.
Thus, this $d\rightarrow\infty$ limit is to be intended at fixed $M$, and the result can be used either to treat generic observables for graphs where the total number of nodes is fixed, or to treat observables that depend only on a finite number of nodes at a time in graphs where the total number of nodes may scale with $d$.

The CLT presented above holds for the variable $\boldsymbol{q}$, and not for the actual distances.
However this is not an issue, as the joint distribution for distances can be derived by a simple coordinate change, factorized over each direction.
Moreover, as we will see in the following, it is often easy to obtain the observables of interest in terms of the $\boldsymbol{q}$ variable.

We now focus on the explicit form of the covariance matrix $\boldsymbol\Sigma$ (notice that, as the vectors $\boldsymbol{q}_{k}$, the covariance matrix is indexed by multi-indices).
By definition, one has
\begin{equation}\label{eq:defSigma}
    \begin{split}
        \boldsymbol{\Sigma}_{(i,j),(k,l)} = \mathbb{E}\left[ (|y_{i}-y_{j}|^{p}-\mu)(|y_{k}-y_{l}|^{p}-\mu) \right]  \, 
    \end{split}
\end{equation}
where $y_{i},y_{j},y_{k},y_{l}$ are all i.i.d. random variables with distribution $\tau$, and $1\leq i<j\leq M$,$\, 1\leq k<l\leq M$.
By permutational symmetry, only three different cases are possible:
\begin{itemize}
    \item \textbf{Diagonal correlations   ($i=k$ and $j=l$)}
        \begin{equation}
            \begin{split}
                \alpha &= \boldsymbol{\Sigma}_{(i,j),(i,j)}\\
                &=
                \int dx dy \, \tau(x)\tau(y) |x-y|^{2p} - \mu^{2}  \, ;
            \end{split}
        \end{equation}
    \item \textbf{Triangular correlations ($i=k$ and $j\neq l$ or  $i\neq k$ and $j=l$)}
        \begin{equation}
            \begin{split}
                \beta &= \boldsymbol{\Sigma}_{(i,j),(i,k)} = \boldsymbol{\Sigma}_{(i,j),(k,j)}\\
                &=\int dx dy dz \, \tau(x)\tau(y)\tau(z) |x-y|^{p}|x-z|^{p} - \mu^{2}  \, ;
            \end{split}
        \end{equation}
    \item \textbf{Pair-pair correlations  ($i,j,k,l$ are all distinct)}
        \begin{equation}
            \begin{split}
                \gamma &= \boldsymbol{\Sigma}_{(i,j),(k,l)}\\
                       &= \left( \int dx dy \, \tau(x)\tau(y) |x-y|^{p} \right)^{2}-\mu^{2}   \, .
            \end{split}
        \end{equation}
        Notice that $\gamma=0$ due to the definition of $\mu$.
\end{itemize}

In the case of the hypercube $\nu = \nu^{\rm cube}$, $\tau(x) = \theta(x)\theta(1-x)$, the coefficients $\alpha$ and $\beta$ are given by:
\begin{equation}
    \begin{split}
        \alpha^{\rm cube} &= \frac{p^{2}(p+5)}{(p+1)^{2}(p+2)^{2}(2p+1)}\\
        \beta^{\rm cube} &= \frac{2}{(p+1)^{2}}\left[ \frac{p^{2}-2}{(2p+3)(p+2)^{2}} + \frac{\Gamma\left( p+2 \right)^{2}}{\Gamma\left( 2p+4 \right)} \right]
    \end{split}
\end{equation}
where $\Gamma(x)$ is the Euler gamma function. In general, $\alpha$ and $\beta$ depend only on the choice of $\tau$.

\begin{figure}
    \centering    
    \scalebox{1.1}{%
        \includegraphics{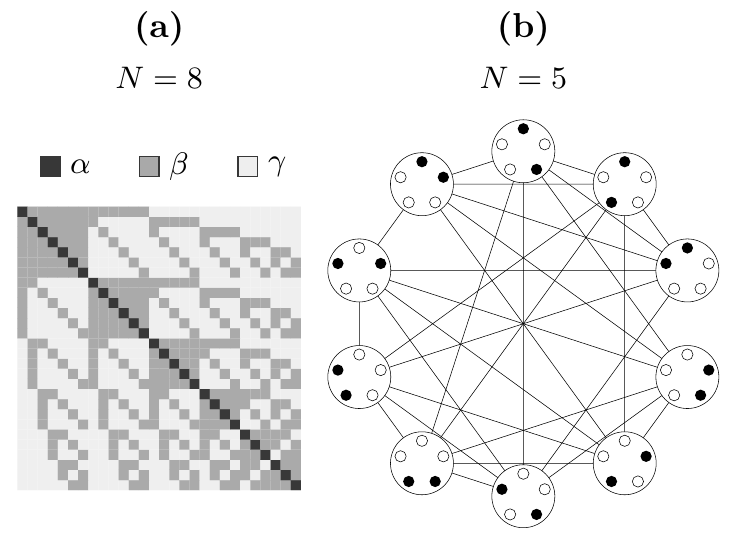}
    }
    \caption{\footnotesize \textbf{(a) Example of a matrix $\boldsymbol{\Delta}(N,\alpha,\beta,\gamma)$ for N=8.}
        The entries with value equal to $\beta$ have the same structure of the adjacency matrix of the Johnson graph.
        \textbf{(b) Example of Johnson graph with N=5.}
        The Johnson graph $J(N,2)$ is the line graph of the complete graph over $N$ nodes.
        It has all the distinct pairs of the original nodes as its vertices, and the vertices are linked if their pairs share and original node.
    }
    \label{fig:Deltamatrix}
\end{figure}


The general form of a matrix with the symmetries of $\boldsymbol\Sigma$ is given by (see Figure~\ref{fig:Deltamatrix}).
\begin{equation} \label{eq:Delta}
    \begin{split} 
        &\boldsymbol{\Delta}_{(i,j)(k,l)}(M,\alpha,\beta,\gamma) \\
        &\qquad=(\alpha -2\beta+\gamma)\delta_{i,k}\delta_{j,l} \\
        &\qquad\quad+
        (\beta-\gamma)(\delta_{i,k}+\delta_{i,l}+\delta_{j,k}+\delta_{j,l}) + \gamma  \, ,
    \end{split}
\end{equation}
where $\delta_{i,j}$ is the Kronecker delta, and $\binom{M}{2}\times\binom{M}{2}$ is the size of the matrix.
We collect properties of such matrices in the following Proposition:
\begin{proposition}
    Let $\boldsymbol{\Delta}$ be a matrix of the form of Equation~\eqref{eq:Delta}, then:
    \begin{enumerate}
        \item it can be written as
            \begin{equation}
                \begin{split}
                    \boldsymbol{\Delta}(M,\alpha,\beta,\gamma) =  (\alpha-\gamma) \boldsymbol{I} + (\beta-\gamma) \boldsymbol{J} + \gamma \boldsymbol{U}  \,
                \end{split}
            \end{equation}
            where $\boldsymbol{I}$ is the identity matrix, $\boldsymbol{U}$ is the matrix with all elements equal to one and $\boldsymbol{J}$ is the adjacency matrix (with null diagonal) of the Johnson graph $J(M,2)$, which is the line graph of the complete graph over $M$ vertices;
        \item the eigenvalues are:
            \begin{itemize}
                \item $\lambda_{1} = \alpha + 2(N-2)\beta + \frac{(N-2)(N-3)}{2} \gamma$ with multiplicity 1;
                \item $\lambda_{2} = \alpha + (N-4)\beta - (N-3) \gamma$ with multiplicity $N-1$;
                \item $\lambda_{3} = \alpha - 2\beta + \gamma$ with multiplicity $\frac{N(N-3)}{2}$;
            \end{itemize}
        \item the inverse matrix $\boldsymbol{\Delta}^{-1}$ is of the same form of $\boldsymbol{\Delta}$ with inverse eigenvalues, and its parameters $\alpha'$, $\beta'$ and $\gamma'$ can be found by solving the linear system
            \begin{equation}
                \begin{split}
                    \lambda_{i}(\boldsymbol{\Delta}(M,\alpha,\beta,\gamma)) \times  \lambda_{i}(\boldsymbol{\Delta}(M,\alpha',\beta',\gamma'))^{-1} =1 \, ,
                \end{split}
            \end{equation}
            for $i=1,2,3$.
    \end{enumerate}
\end{proposition}
\pf
\textbf{1)} Follows from the explicit expression of $\boldsymbol{I},\boldsymbol{J}$ and $\boldsymbol{U}$.\\
\textbf{2)} $\boldsymbol{I},\boldsymbol{J}$ and $\boldsymbol{U}$ commute between each other, and can be diagonalized simultaneously. The contribution of $\boldsymbol{I}$ is trivial. $\boldsymbol{J}$ and $\boldsymbol{U}$ share a non-degenerate eigenvector (that with all components equal to one), that accounts for $\lambda_{1}$. In the orthogonal subspace, $\boldsymbol{U}$ represents the null operator, and does not contribute. Thus, the remainder of the spectrum is determined by that of $\boldsymbol{J}$, which is known \cite{burcroff2017johnson}. \\
\textbf{3)} Follows from the fact that a matrix and its inverse share the same eigenvectors.
\qed

\section{Number of cliques reveals non-trivial structure of hard geometric graphs}
\label{sec:res}

We are now ready to compute observables on random geometric graphs in the limit of infinite dimensions; in particular, we aim to characterize the average number of subgraph with a given structure.
Recall that the \emph{adjacency matrix} of a graph $g$ with $M$ nodes is the $M\times M$ matrix with entry $A_{ij}(g)=1$ if $(i,j)$ is an edge of $g$, and $A(g)_{ij}=0$ otherwise.

In general, the average number of a certain subgraph $g$ with $M$ nodes of a random geometric graph can be factored in two terms.
The first one is a combinatorial factor $\binom{N}{M}$, that accounts for the number of ways in which one can extract $M$ nodes from a set of $N$ of them.
The second one is the so-called density $\rho_{g}(r)$ of the subgraph $g$ at scale $r$, that is the probability that $M$ random points are close enough with respect to the cutoff radius $r$ to form a subgraph with the same adjacency matrix of $g$.
Recalling the definition of the joint probability of the distances between $M$ points given in Equation~\eqref{eq:PIdist}, we have that
\begin{equation}
    \begin{split}
        \rho_{g}(r) 
        &= \int d\boldsymbol{y} \, \Pi(\boldsymbol{y}) \prod_{1 \leq i < j \leq M} \left[ h_{r}\left( y_{(i,j)} \right) \right]^{A_{ij}(g)} \, ,
    \end{split}
\end{equation}
where $y_{(i,j)}$ is the distance between nodes $i$ and $j$.
We can rescale the variables $y_{(i,j)}$ as in Equation~\eqref{eq:defQ}, and exploit the fact that for large dimension $d\boldsymbol{y} \, \Pi(\boldsymbol{y}) \sim d\boldsymbol{q} \,  \mathcal{N}(\boldsymbol{0},\boldsymbol{\Sigma})(\boldsymbol{q})$ (see Theorem~\ref{thm:CLT}) to obtain an expression for $\rho_{g}(r)$ that is valid in the limit of large dimension:
\begin{equation}
    \begin{split}
        \rho_{g}(r) 
        &= \int d\boldsymbol{y} \, \Pi(\boldsymbol{y}) \prod_{1 \leq i < j \leq M} \left[ h_{r}\left( y_{(i,j)} \right) \right]^{A_{ij}(g)}\\
        &\sim  \int d\boldsymbol{q} \, \mathcal{N}(\boldsymbol{0},\boldsymbol{\Sigma})(\boldsymbol{q}) \\
        &\quad\prod_{1 \leq i < j \leq M} \left[ h_{r}\left( \left[d \, \mu + \sqrt{d} \, q_{(i,j)}\right]^{\min\left(1,\frac{1}{p}\right)} \right) \right]^{A_{ij}(g)}
        \, 
    \end{split}
\end{equation}
where $\mathcal{N}(\boldsymbol{0},\boldsymbol{\Sigma})$ is the multivariate Gaussian with null mean and covariance $\boldsymbol\Sigma$ (given in Equation~\eqref{eq:defSigma}), i.e.
\begin{equation}
    \begin{split}
    \mathcal{N}(\boldsymbol{0},\boldsymbol{\Sigma})(\boldsymbol{q}) = \frac{e^{ - \frac{1}{2} \boldsymbol{q}^{T} \boldsymbol{\Sigma} \boldsymbol{q} }}{\sqrt{(2\pi)^{\binom{M}{2}}  \det \boldsymbol{\Sigma}}}
    \, .
    \end{split}
\end{equation}
In the rest of the section, all results are to be intended in the limit of large dimension. 

As a paradigmatic example, we consider the average density of $M$-cliques $\rho_{M}(r)$, i.e. fully connected subgraphs with $M$ vertices, on random geometric graphs with generic activation function $h_{r}(x)$; in this specific case, $A_{ij}$ has only unit entries, so that
\begin{equation}
    \begin{split}
        \rho_{M}(r) 
        &= \int d\boldsymbol{q} \, \mathcal{N}(\boldsymbol{0},\boldsymbol{\Sigma})(\boldsymbol{q}) \\
        &\quad\prod_{1 \leq i < j \leq M} h_{r}\left( \left[d \, \mu + \sqrt{d} \, q_{(i,j)}\right]^{\min\left(1,\frac{1}{p}\right)} \right)
        \, .
    \end{split}
\end{equation}

In the case of hard activation function $h^{\rm hard}$, we observe that
\begin{equation}
    \begin{split}
        h^{\rm hard}_{r}(x) &= h^{\rm hard}_{r^{p}}(x^{p}) \\
        h^{\rm hard}_{r}(x+c) &= h^{\rm hard}_{r-c}(x) \, , \quad \forall c \in \mathbb{R}\\
        h^{\rm hard}_{r}(x) &= h^{\rm hard}_{c\, r}(c \, x) \, , \quad \forall c \in \mathbb{R}^{+}
         \, 
    \end{split}
\end{equation}
so that the $p$-th root can be discarded along with a factor of $\sqrt{d}$, and the integral reduces to 
\begin{equation}
    \begin{split}
        \rho^{\rm hard}_{M}(r) = \overline{\rho}^{\rm hard}_{M}\left( \frac{r^{\max(1,p)}-d\,\mu}{\sqrt{d}} \right) 
        \, ,
    \end{split}
    \label{rhofactor}
\end{equation}
with
\begin{equation} \label{eq:HardRHOteo}
    \begin{split}
        \overline{\rho}^{\rm hard}_{M}(x)  &= 
        \int d\boldsymbol{q} \, \mathcal{N}(\boldsymbol{0},\boldsymbol{\Sigma})(\boldsymbol{q}) \prod_{1 \leq i < j \leq M} h^{\rm hard}_{x}\left( q_{i,j} \right)\\
                                           &= \int d\boldsymbol{q} \, \mathcal{N}(\boldsymbol{0},\boldsymbol{\Sigma})(\boldsymbol{q}) \prod_{1 \leq i < j \leq M} \theta\left( x - q_{(i,j)} \right) 
    \end{split}
\end{equation}
which is a multivariate Gaussian cumulative distribution function. Equation~\eqref{rhofactor} highlights the simple dependence of $\overline{\rho}^{\rm hard}_{M}$ on the parameters $p, d$ and $\mu$.

In the case $M=2$, the integral in Equation~\eqref{eq:HardRHOteo} can be explicitly solved as it reduces to the computation of an error function, giving
\begin{equation} \label{eq:HardRHOteo2}
    \begin{split}
        \overline{\rho}^{\rm hard}_{2}(x) = \frac{1}{2}\left[1 + \text{Erf} \left( \frac{x}{\sqrt{2\alpha}} \right) \right]
        \, .
    \end{split}
\end{equation}

\begin{figure*}[t]
    \centering    
    \scalebox{1}{%
        \includegraphics{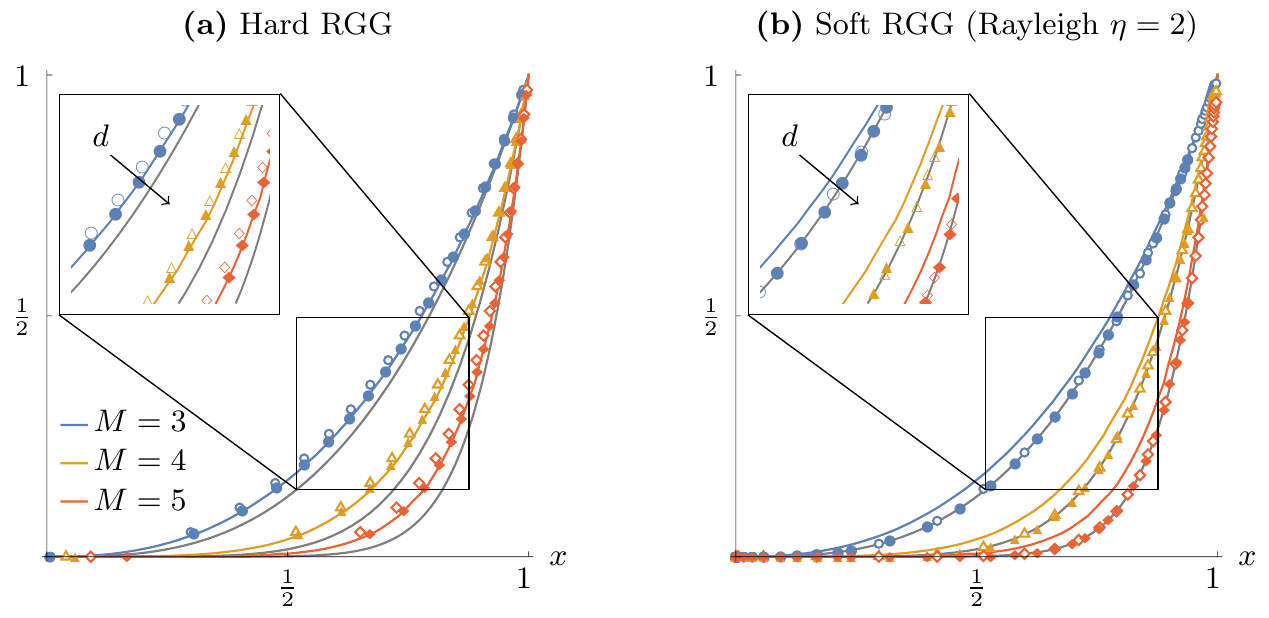}
    }
    \caption{\footnotesize \textbf{Comparison between finite $d$ simulations and infinite $d$ analytical predictions.}
        Colored solid lines represent the analytical predictions for $\omega^{\rm hard}_{M}(x)$ obtained from Equation~\eqref{eq:HardRHOteo}, for $M=3,4,5$ [blue (first line from above), orange (third line from above) and red (fifth line from above) respectively]. 
        Gray solid lines represent $\omega^{\rm ER}_{M}(x)$ for the Erd\"os-R\'enyi graph [Equation~\eqref{eq:omegaER}] for comparison for the same values of $M=3,4,5$ (second, fourth and sixth line from above respectively).
        Open and filled markers show numerical simulations at $d=20,200$ respectively, $p=2$ and $\nu=\nu^{\rm cube}$, for hard RGG \textbf{(a)} and soft RGG with Rayleigh activation function $\eta=2$ \textbf{(b)}, for the same values of $M=3,4,5$ (circles, triangles and diamonds respectively).
        In practice, $\omega_{M}(x)$ can be represented by producing a scatter plot of $\rho_M(r)$ versus $\rho_2(r)$.
    }
    \label{fig:Numerics}
\end{figure*}

The simple dependence of $\rho_{M}(r)$ on $p,d$ and $\mu$ suggests to study the quantities 
\begin{equation}
    \begin{split}
        \omega^{\rm hard}_{M}(x) &=
(\rho^{\rm hard}_{M} \circ (\rho^{\rm hard}_{2})^{-1})(x) \\
                                 &=
(\overline{\rho}^{\rm hard}_{M} \circ (\overline{\rho}^{\rm hard}_{2})^{-1})(x)
\, .
    \end{split}
\end{equation}
    Notice that $\omega_{M}(x)$ is related to $\rho_{M}(r)$ by the bijective change of variable $x = \rho_{2}(r)$ that, to a cutoff radius $r$, assigns the probability $x$ that a random pair of nodes in the graph will be linked.
Thus, $\omega_{M}(x)$ gives the probability that $M$ random nodes will form a $M$-clique as a function of the probability that two random nodes will be linked.
In practice, $\omega_{M}(x)$ can be plotted by producing a scatter plot of $\rho_M(r)$ versus $\rho_2(r)$.
With this change of variable, the dependence of $\rho_{M}$ on $p,d$ and $\mu$ cancels out, and the curves at different values of the parameters all lie in the domain $x\in[0,1]$. 

In the case of soft random geometric graphs with continuous activation functions, one can expand $h_{r}(x)$ to the $0$-th order in powers of $1/\sqrt{d}$, obtaining that in the limit of high dimension
\begin{equation}
    \begin{split}
        \rho^{\rm regular}_{M}(r) &= \left[\rho^{\rm regular}_{2}(r)\right]^{\binom{M}{2}} \\
        \rho^{\rm regular}_{2}(r) &= h_{r}\left( (d \mu)^{\min\left( 1,\frac{1}{p} \right) } \right)  \\
    \end{split}
\end{equation}
Here, the relation between $\rho_{M}$ and $\rho_{2}$ reduces to that of Erd\"os-R\'enyi graphs with linking probability $\rho^{\rm regular}_{2}(r)$, i.e. 
\begin{equation} \label{eq:omegaER}
    \begin{split}
        \omega^{\rm soft}_{M}(x)=\omega^{\rm ER}_{M}(x) = x^{\binom{M}{2}}
         \, .
    \end{split}
\end{equation}

In the special case of Rayleigh fading activation function $h^{\rm rayleigh}$, one has
\begin{equation}
    \begin{split}
        \rho^{\rm rayleigh}_{2}(r) = \exp\left[ -\xi \left( \frac{d \mu}{r} \right)^{\eta\min\left( 1,\frac{1}{p} \right) }  \right] \, .
    \end{split}
\end{equation}

Intuitively, the difference between hard and soft RGGs depends on the freedom in performing the rescaling of the cutoff radius in the former case [see Equation~\eqref{rhofactor}], which is lost in the latter.

We performed extensive numerical simulations to study Equation~\eqref{eq:HardRHOteo} and to check our analytical results; a summary is provided in Figure~\ref{fig:Numerics}. The numerical methods are described in Appendix~\ref{app:numerics}. 
We observe a very good qualitative agreement between our analytical predictions in infinite dimension and finite dimensional simulations for both hard ans soft random geometric graphs.
    The convergence to the limit is fast, and even at $d=20$ our analytical prediction provides a good approximation of the simulated observables.
    More quantitatively, we observe relative deviations from the analytical predictions of the order of $\sim 10\%$ in $d=20$ and $\sim 2\%$ in $d=200$ in both the hard and soft case for $k=3$.
    For $k=4,5$ relative errors are slightly larger, mainly due to the fact the we are measuring $\rho_{k}(r)$ by a random sampling procedure (see Appendix~\ref{app:numerics}) that needs more and more samples as $k$ increases.

\section{Discussion}
\label{sec:discuss}

In this work we exploited a multivariate version of the central limit theorem to compute average observables of random geometric graphs in the limit of infinite dimension. In particular, we obtained the average number of $M$-cliques in hard and soft RGGs for different distance functions induced by $p$-norms.

Our approach highlights that convergence to the ERG prediction for local observables depends on the choice of the ensemble: soft RGGs in particular seem to approach this naive limit for $d\rightarrow \infty$, whereas hard RGGs whose probability distribution of the nodes fulfils the CLT hypothesis deviate systematically from it. This result suggests that the latter provide a non-trivial null model to benchmark empirical data.

A potentially useful application of our results lies in their guidance with regards to the choice of \emph{null models},
which are essential if one is to extract meaningful information from the data.
For example, let us consider data points from an empirical data set (such as MNIST, for instance), 
and a graph constructed on these points, where a link exist whenever two data points are closer than a given cutoff radius 
(determining this graph is the starting point of algorithms for hierarchical clustering or manifold learning).
Now, say the number of cliques in this graph deviates from the ER prediction.
If we erroneously believe that RGGs in high dimension are ERGs, then we should conclude that the behavior
is due to specificities of the data (e.g., deviations from the assumption of independence).
This conclusion would be misleading, since, for the hard activation function,
there are systematic deviations from the ER prediction even if the data points are uncorrelated and identically distributed.
Our work makes clear that ruling out the null hypothesis of RGG in high dimension is fundamentally different from ruling out the hypothesis of being a ERG.

Since the CLT can be formulated in a much more general setting than the one reported in this manuscript, we expect that our findings hold (possibly with slight modifications) for several probability distributions of the nodes not included here, e.g. not factorized over coordinates, but with mild inter-coordinate correlations; factorized over coordinates, but not identically distributed; factorized over coordinates, but with infinite second moment.
The wide basin of attraction of the Gaussian limit hints to the possibility that the properties of high-dimensional structured datasets may be faithfully described by our approach.
In this manuscript we worked with the simplest version of the CLT, as random geometric graphs are commonly studied with nodes that are independently drawn in the hypercube.
    The very relevant case of \emph{structured data} \cite{Borra_2019,rotondo2019CountingLearnableFunctions,pastore2020statistical,rotondo2020beyond} calls for more sophisticated CLTs, which can be addressed with the same tools developed here.

Another potentially interesting case is that of RGGs whose vertex measure is supported on low-dimensional manifolds but is embedded in a much higher-dimensional ambient space with noise.
Which observables will be hidden by the added noise? And which will be robust, allowing to recover non-trivial properties of the underlying geometry?

Finally, our numerical simulations show that the infinite dimensional limit is a good approximation even in finite dimensions of order $d\sim10$. This hints at the possibility to improve our results by computing higher order corrections to the CLT, and using $d$ as a perturbative parameter, to access the low dimensional regime of RGGs.

\section*{Acknowledgments}
P. R. acknowledges funding by the INFN Fellini program H2020-MSCA-COFUND-2016 Grant Agreement n. 754496.



\appendix

\section{Numerical methods}
\label{app:numerics}

To numerically evaluate the integrals of Equation~\eqref{eq:HardRHOteo}, we implemented the algorithm described in \cite{genz1992NumericalComputationMultivariate}, allowing very fast run times for the small values of $M$ ($M\lesssim10$) we where interested in; notice that the dimension of the integral is already of order $10^{2}$ for $M=10$.
Higher values of $M$ would require finer techniques.

To compute the density of $M$-cliques in simulated hard RGGs, we implemented a simple random sampling procedure, as exhaustive enumeration scales poorly, i.e. as $O\left(N^{M}\right)$, with the total number of nodes.
For each realization of the nodes (with $\nu^{\rm cube}$ and $N=10^{4}$), we extracted $\sim 5 \cdot 10^{5}$ $M$-uples of nodes, computing the minimum cutoff distance at which they formed a clique.
The cumulative distribution of the minimal distances obtained, averaged over different realization of the nodes, reconstructs $\rho^{\rm hard}_{M}(r)$.
We noticed that as $N$ grows, the last average is well approximated by a single realization of the nodes, suggesting a self-averaging property for the density of $M$-cliques; in practice, not averaging does not affect the results of the simulations.

To compute the density of cliques in simulated soft RGGs with generic activation function, we implemented again a random sampling procedure.
This time, for each realization of the nodes (as above) and for a fixed radius $r$, we counted how many of $\sim 10^{4}$ $M$-uples of nodes $\{y_{i}\}_{i=1}^{M}$ where $M$-cliques, considering each of them to be a $M$-clique with probability
\begin{equation}
    \begin{split}
        \prod_{1\leq i < j \leq M} h_{r}(d(\vec{y}_{i},\vec{y}_{j}))\, .
    \end{split}
\end{equation}
Normalizing the count over the total number of candidate cliques and averaging over different realizations of the nodes (order $~10^{2}$) gives an empirical estimation for $\rho_{M}(r)$ in the soft case.

\end{document}